\documentclass[twocolumn]{aanomarks}  
\usepackage{graphicx}
\usepackage{natbib}
\usepackage{color}           
\definecolor{mrk}{RGB}{0,0,0}
\newcommand{\mrk}[1]{{\color{mrk} #1}}
\definecolor{dltd}{RGB}{200,100,100}

\newcommand{\ang}{\mathrm{\AA}}

\begin{document}

\title{Analysis of unresolved photospheric magnetic field structure using Fe~I 6301 and 6302 lines}

   \author{M. Gordovskyy \inst{1}\fnmsep\thanks{\email{mykola.gordovskyy@manchester.ac.uk}}, 
	  S. Shelyag\inst{2}, 
	  P.K. Browning\inst{1}   
          \and
	V.G. Lozitsky\inst{3}}

   \institute{Jodrell Bank Centre for Astrophysics, University of Manchester, Manchester M13\,9PL, UK
          \and
Department of Mathematics, Physics and Electrical Engineering, University of Northumbria, Newcastle NE1\,8ET, UK
         \and
             Astronomical Observatory, Kyiv National University, Observatorna 3, Kyiv 01053, Ukraine
             }

   \date{Received ; accepted }

\authorrunning{Gordovskyy et al.}
\titlerunning{Unresolved photospheric magnetic field}

 
  \abstract
   {Early magnetographic observations indicated that magnetic field in the solar photosphere has unresolved small-scale structure. Near-infrared and optical data with extremely high spatial resolution show that these structures have scales of few tens of kilometres, which are not resolved in the majority of solar observations.}
   {The goal of this study is to establish the effect of unresolved photospheric magnetic field structure on Stokes profiles observed with relatively low spatial resolution. Ultimately, we aim to \mrk{develop methods for fast estimation of the photospheric magnetic filling factor and line-of-sight gradient of the photospheric magnetic field, which can be applied to large observational data sets}.}
   {We exploit 3D MHD models of magneto-convection developed using MURAM code. Corresponding profiles of Fe~I 6301.5 and 6302.5~$\ang$ spectral lines are calculated using NICOLE radiative transfer code. The resulting I and V Stokes [x,y,$\lambda$] cubes with reduced spatial resolution of 150~km are used to calculate magnetic field values as they would be obtained in observations with Hinode/SOT or SDO/HMI. }
   {Three different methods of the magnetic filling factor estimation are considered: the magnetic line ratio method, Stokes V width method and a simple statistical method. We find that the statistical method and the Stokes V width method are sufficiently reliable for fast filling factor estimations. Furthermore, we find that Stokes $I\pm V$ bisector splitting gradient can be used for fast estimation of line-of-sight gradient of the photospheric magnetic field.}
   {}


   \maketitle
%

\section{Introduction}\label{s-intro}

\begin{figure*}[ht!]    
\centerline{\includegraphics[width=0.8\textwidth,clip=]{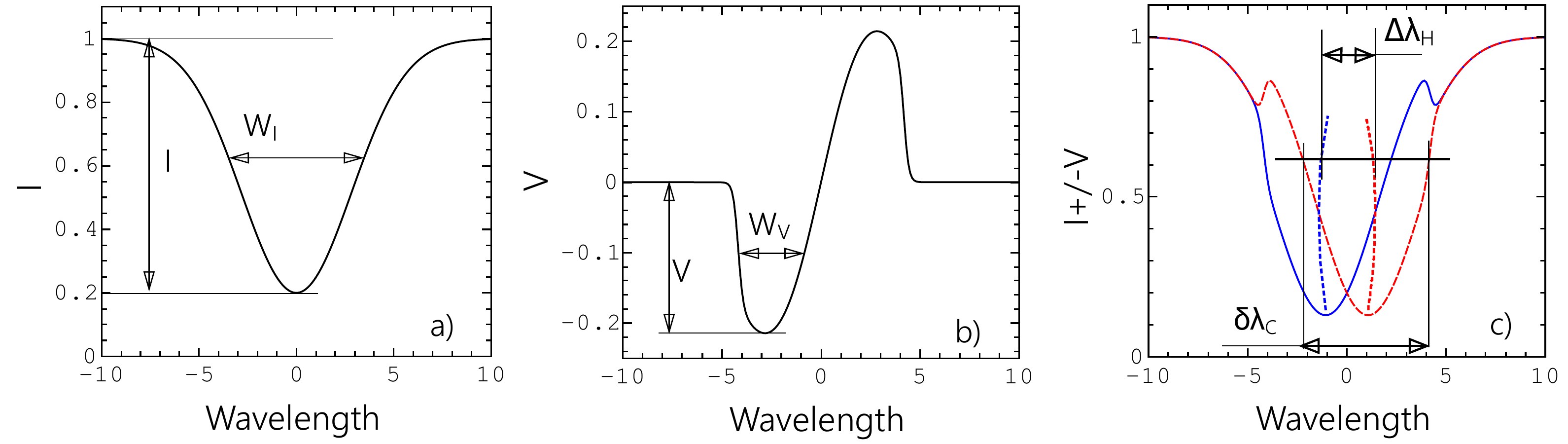}}
\caption{Sketch of a line profile, showing Stokes I profile (panel a), Stokes V (panel b), and Stokes I$\pm$V profiles (panel c) with some of their parameters used in this paper.}
\label{f-sketch}
\end{figure*}

\begin{figure*}[ht!]    
\centerline{\includegraphics[width=0.7\textwidth,clip=]{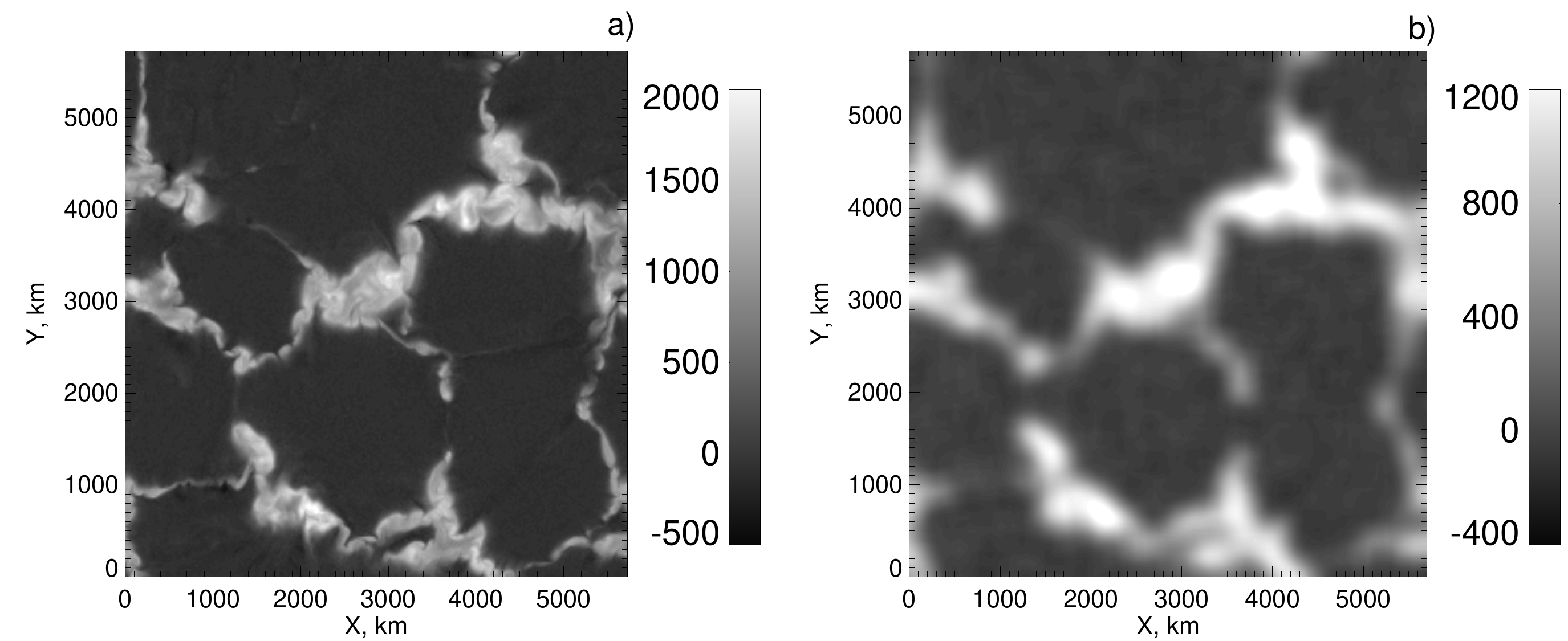}}
\caption{Magnetic field in snapshot C at different spatial resolutions. Panel (a): Magnetic field measured using Stokes V amplitudes of the 6301 line with high spatial resolution (not calibrated). Panel (b): Map of B$_\mathrm{eff}$[6301], magnetic field from the same model snapshot obtained using degraded Stokes profiles, representing low-resolution data.}
\label{f-fieldmap}
\end{figure*}

Early observational studies indicated that photospheric magnetic field is very non-uniform on scales smaller than the spatial resolution of optical instruments. Magnetographic observations showed that the magnetic field strengths measured using two Fraunhofer lines with similar characteristics but different magnetic sensitivity ({\it i.e.} different Lande factor $g$) can differ by factor of up to $2.5$ \citep{host72,sten73}. This has been interpreted as evidence of strong horizontal field inhomogeneity. It has been suggested that most of the photospheric magnetic flux is carried by numerous intense small-scale magnetic fluxtubes, and the photospheric magnetic filling factor can be as low as 10\% \citep{frst72}. Despite a very significant progress in achieving high spatial resolution in solar observations, most existing optical solar instruments do not resolve the smallest scales in the photospheric magnetic field \cite[see e.g.][for review]{sola93,lolo94,sosc04,dewe08}. 

Obviously, direct high-resolution observations are the most reliable way of studying photospheric fine structure. However, very high spatial resolution (50-100~km or even less) can be achieved only in some observations using advanced instrumentation as well as advanced data-processing techniques, which are often computationally expensive. Direct high-resolution observations of small-scale photospheric magnetic elements were performed using speckle-interferometry in Fe~I~5250.2~$\ang$ line \citep{keva92,kell92}. They found magnetic elements with a field strength of a few kG and estimated their sizes at $100\,-\,200$~km. \citet{lin95} observed Stokes profiles in magneto-sensitive near-infrared Fe~I lines 15648~$\ang$ and 15652~$\ang$ and showed that there are two types of small-scale 
magnetic elements: stronger elements with the field of $1.4$~kG and diameters 100--1000~km located in the network boundaries, and weaker ones with fields of about $ 500$~G and diameters about $70$~km, located inside granulation cells. \citet{lage10} using IMaX magnetograph on-board Sunrise balloon mission have achieved spatial resolution of about 100~km in all Stokes components. They were able to detect small magnetic elements with the filling factor equal 1 (i.e. no unresolved structure), which have been interpreted as individual photospheric fluxtubes. The sizes of these elements are 100-500~km.

Inversion of Stokes profiles observed using moderate spatial resolution (few 100~km) can be an alternative to high-resolution observations \cite[e.g.][]{vite11}. However, inversion algorithms are computationally expensive and can be applied only to rather small patches of the photosphere. Furthermore, most inversion codes oversimplify the magnetic field structure by assuming the same magnetic filling factor at different heights. 

Forward-modelling is another way of studying small-scale photospheric magnetic field. 3D models of magneto-convection in the photosphere developed using high-resolution 3D magnetohydrodynamic (MHD) simulations offer a unique opportunity to investigate photospheric magnetic field structure unresolved by normal solar telescopes. Combined with the radiative transfer calculations, these models make it possible to link characteristics of small- and large-scale magnetic field 3D structure with parameters inferred directly from solar observations with limited spatial and spectral resolution.

Recently, \citet{smso17} used MHD models of magneto-convection combined with a radiative transfer calculations in order to investigate Stokes profiles of several photospheric lines. The focus was on using \mrk{various} pairs of lines for measuring photospheric field using the magnetic line ratio (MLR) approach. This approach makes it possible to evaluate actual field value based on the ratio of magnetic field values measured using two lines with similar thermodynamic characteristics (and, hence, similar formation depths) but different Lande factors, unlike single line measurements, which yield average field (or magnetic flux) values (see Section~\ref{s-callibr}). This study identified two new pairs of lines, one visible and another in the near-infrared, as effective diagnostic tools for magnetic field measurements using MLR approach.  

In the present study, \mrk{we investigate the effect of magnetic field filling factor (in a plane perpendicular to the LOS) and vertical field gradient on Stokes I and V profiles observed with relatively low spatial resolution.} We deploy techniques very similar to \citet{smso17}, however, we focus specifically on unresolved structure of photospheric field, both in horizontal and vertical (i.e. line-of-sight, LOS) directions. \mrk{Most importantly, in addition to the magnetic line ratio method, we consider two other methods (Stokes V width method and the statistical method).}  The ultimate goal is to find a simple, empirical way of estimating these two parameters using large field-of-view spectropolarimetric data from telescopes such as Hinode, Gregor and future DKIST.

\section{Methodology and definitions}

\begin{table}
\begin{center}
\begin{tabular}{ l l l }
\hline
   & Resolution, km & Pixel size, km \\
\hline
Hinode / SOT & 190 & 180 \\
Gregor / GFPI & 63 & 26 \\
DKIST / ViSP & 50 & ? \\
Synthetic data & 100 & 50 \\
\hline
\end{tabular}
\end{center}
\caption{Spatial resolution around 6300$\ang$ and pixel size for three optical solar imaging spectrographs and for the synthetic degraded Stokes cubes used in this study \citep{kose07,vole10,trie16}.}
\end{table}

\begin{figure*}[ht!]    
\centerline{\includegraphics[width=0.85\textwidth,clip=]{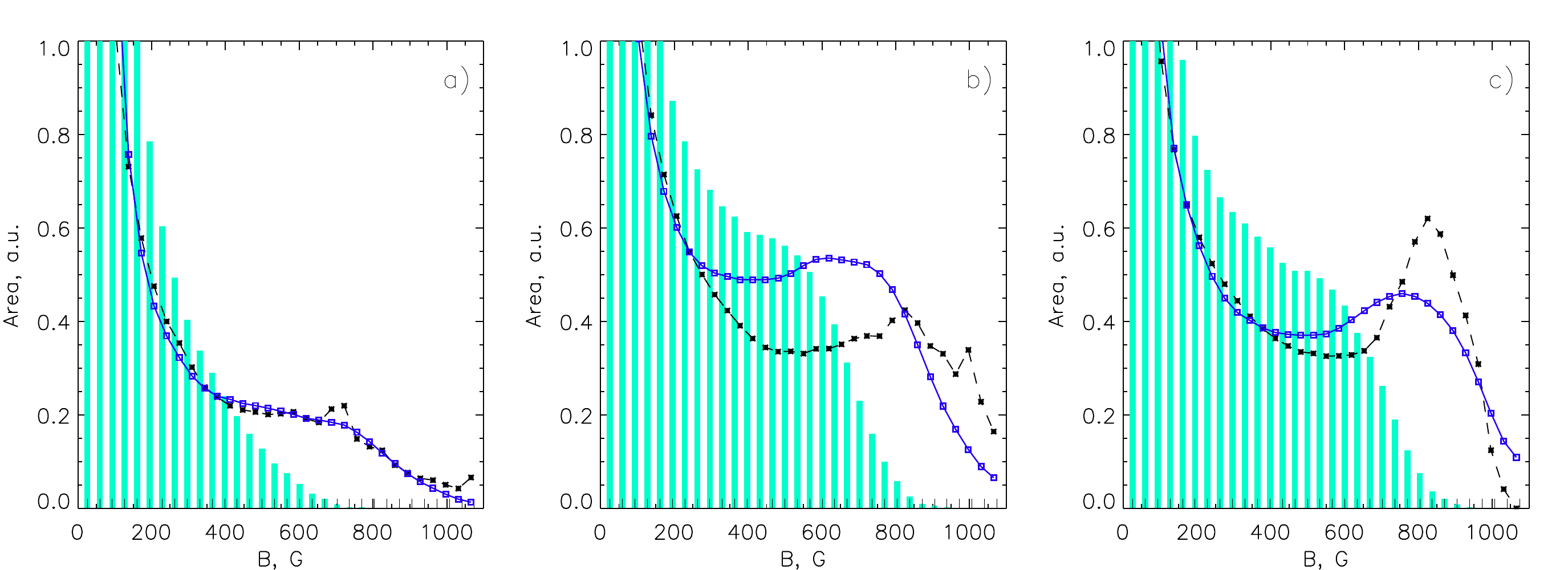}}
\caption{Magnetic field distribution function in snapshots A, B and C (panels a, b and c, respectively). The black dashed lines correspond to the actual field taken from the high-resolution MHD simulations and averaged along LOS along 400~km below the temperature minimum. Solid blue lines with squares correspond to the magnetic field measured with high resolutions using Stokes V amplitudes of 6301 line. Blue bars correspond to the effective magnetic field  B$_\mathrm{eff}$[6301], {\it i.e.} magnetic field values measured with Stokes V amplitudes of 6301 line using spatially-degraded data (see Section~\ref{s-low}).}
\label{f-histomag}
\end{figure*}
\begin{figure*}[ht!]    
\centerline{\includegraphics[width=0.85\textwidth,clip=]{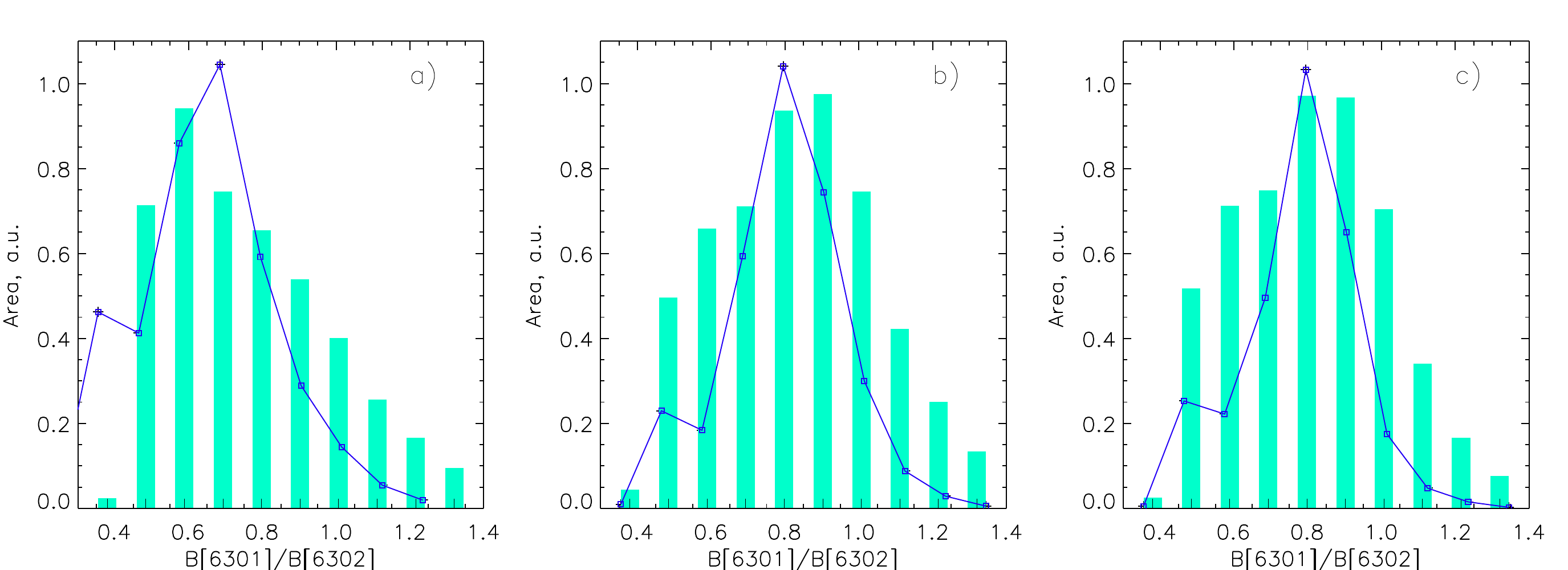}}
\caption{MLR value spectra from snapshot snapshots A, B and C (panels a, b and c, respectively). The line corresponds to MLR measured with original MHD model resolution, while bars correspond to MLR measured using Stokes V amplitudes obtained with low spatial resolution.}
\label{f-historat}
\end{figure*}
\begin{figure*}[ht!]    
\centerline{\includegraphics[width=0.85\textwidth,clip=]{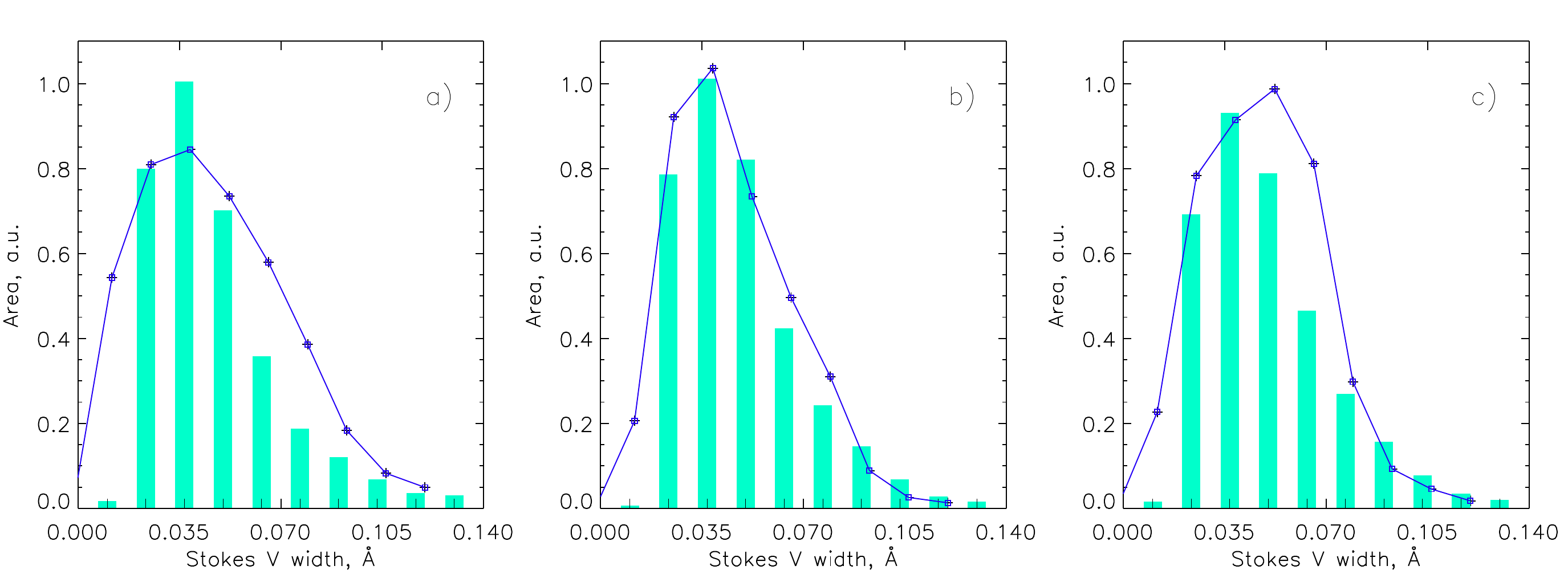}}
\caption{The same as in Figure~\ref{f-historat} but for the Stokes V width values.}
\label{f-histowid}
\end{figure*}

In this paper, we use a combination of MHD and radiative transfer simulations in order to investigate Stokes profiles of Fe~I 6301.5 and 6302.5~$\ang$ lines produced by the photospheric magnetic field. Comparison of synthetic Stokes profiles calculated with high and low spatial resolution for the same areas of the model photosphere make it possible to link characteristics of the low-resolution Stokes profiles with intrinsic physical characteristics of photospheric magnetic field. 

\mrk{The 6301 - 6302~$\ang$ is not the best pair for the magnetic line ratio method \cite[see analysis in][]{khoe05,khco07,smso17}. However, the 6301 - 6302 pair is currently the most used and most widely-available pair, thanks to Hinode SOT and a few other instruments, and that is why this specific pair is chosen. 

Both lines are formed in a relatively large height interval (400-500~km), however, the average formation height of the 6301~$\ang$ line is 80-150~km higher than that of the 6302~$\ang$ line (depending on thermodynamic conditions). This difference is substantial compared to other pairs: the classical 5247 - 5250~$\ang$ pair has formation height difference of about 50~km, while for the near-infrared pair 15648 - 15652~$\ang$ it is less than 20~km \cite[see e.g. Figure 5 in][]{khco07}. At the same time, the difference in the formation heights of 6301 and 6302 lines is substantially smaller than the height intervals over which the lines are formed and, hence, this pair can still be used in MLR method, although one should expect relatively large errors.}

By ``high-resolution data'' we mean the original data from MHD simulations, with has horizontal spatial resolution of 6.25~km (see Section~\ref{s-mhd}). By ``low resolution data'' we mean data degraded to spatial resolution of about 150~km, which is typical for most optical solar observations (see Table 1), and by ``effective'' we mean parameters obtained using this low-resolution data (see Section~\ref{s-low}). By ``small-scale'' or ``sub-telescopic'' structures we mean structures with the length-scales smaller than $\sim$100~km, {\it i.e.} those that are resolved in the high-resolution data, but unresolved in the low-resolution data. \mrk{By intrinsic magnetic field we mean value of the vertical component of magnetic field $B_z$ averaged in the height interval between -100 and 300~km.}

We assume that the observed photospheric areas are close to the disk centre and, therefore, by vertical or line-of-sight (LOS) direction we mean z-direction in the MHD model. The velocity sign is defined as in spectroscopy: positive velocities correspond to the motion away from the observer ({\it i.e.} downflows, Doppler shift to longer wavelengths), and negative velocities correspond to the motion towards the observer ({\it i.e.} upflows, Doppler shift to shorter wavelengths).

In this paper we use standard definitions of commonly used parameters, unless stated otherwise; some of them are shown in the line profile sketch in Figure~\ref{f-sketch}. The amplitude and the full width at half-maximum (FWHM) of Stokes I profile are denoted as $I$ and $W_I$. However, it should be noted that by line width we mean FWHM of I$\pm$V, otherwise stated. The amplitude and width of Stokes V profile are the amplitude and FWHM of the blue peak of line's Stokes V profile, they are denoted $V$ and $W_V$, respectively. The bisector splitting function is defined as $\Delta \lambda_H (\delta \lambda_C)$, where $\Delta \lambda_H$ is the splitting of the Stokes I+V and I-V bisectors at a certain intensity level, and $\delta \lambda_C$ is the full width of the profile at that intensity level. The bisector splitting gradient BSG is defined as BSG$=\frac{d(\Delta \lambda_H) }{d(\delta \lambda_C)}$. 
The MLR is defined as the ratio of Stokes V amplitude of the line with higher Lande factor to that of the line with lower Lande factor. Namely, MLR$ = \frac{V[6301]}{V[6302]}$, where $V[6301]$ and $V[6302]$ are amplitudes of the blue peaks of the Stokes V profiles of these lines.

\mrk{In order to measure the amplitude and width of the blue wing of Stokes V profile, we locate positive and negative extrema for each line. The extremum located at shorter wavelength is assumed to be the blue peak; its amplitude and full width at half-maximum are $V[630x]$ and $W_V[630x]$, respectively.}

Effective values (i.e. those corresponding to the low resolution) of magnetic field, velocity, Stokes I and V profile width are obtained using spatially-averaged Stokes profiles. Degraded Stokes cubes I$_\mathrm{D}$(x',y',$\lambda$) and 
V$_\mathrm{D}$(x',y',$\lambda$) are calculated by convolving the original Stokes cubes with a 2D Gaussian PSF and adding random noise as follows: 

\[
S_\mathrm{P}(x,y,\lambda) = \frac 1A S(x,y,\lambda) * \exp \left( - 2.77 \frac {x^2 + y^2}{r_\mathrm{PSF}^2} \right) ,
\]
\[
S_\mathrm{D}(x',y',\lambda) = \xi (x',y') \frac {\int \limits_{-\beta}^{+\beta} \int \limits_{-\beta}^{+\beta} xy S_\mathrm{P}(x'+x,y'+y,\lambda) dx dy}{\int  \limits_{-\beta}^{+\beta} \int \limits_{-\beta}^{+\beta} xy dx dy},
\]
where $S(x,y,\lambda)$ is the original and $S_\mathrm{D}(x,y,\lambda)$ is degraded Stokes ($I$ or $V$) profiles; respectively. $(x,y)$ and $(x',y')$ are original grid and sparse grid coordinates, respectively, and $\beta = r_\mathrm{PSF}/4$. Here $r_\mathrm{PSF}=100$~km is, effectively, the spatial resolution, and the step of the sparse grid is $r_\mathrm{PSF} /2$. The function $\xi (x',y')$ adds random noise with normal distribution and $\sigma=$10$^{-3}$.

\section{Magnetic field structure obtained using magneto-convection simulations}\label{s-mhd}

\begin{figure*}[ht!]    
\centerline{\includegraphics[width=0.7\textwidth,clip=]{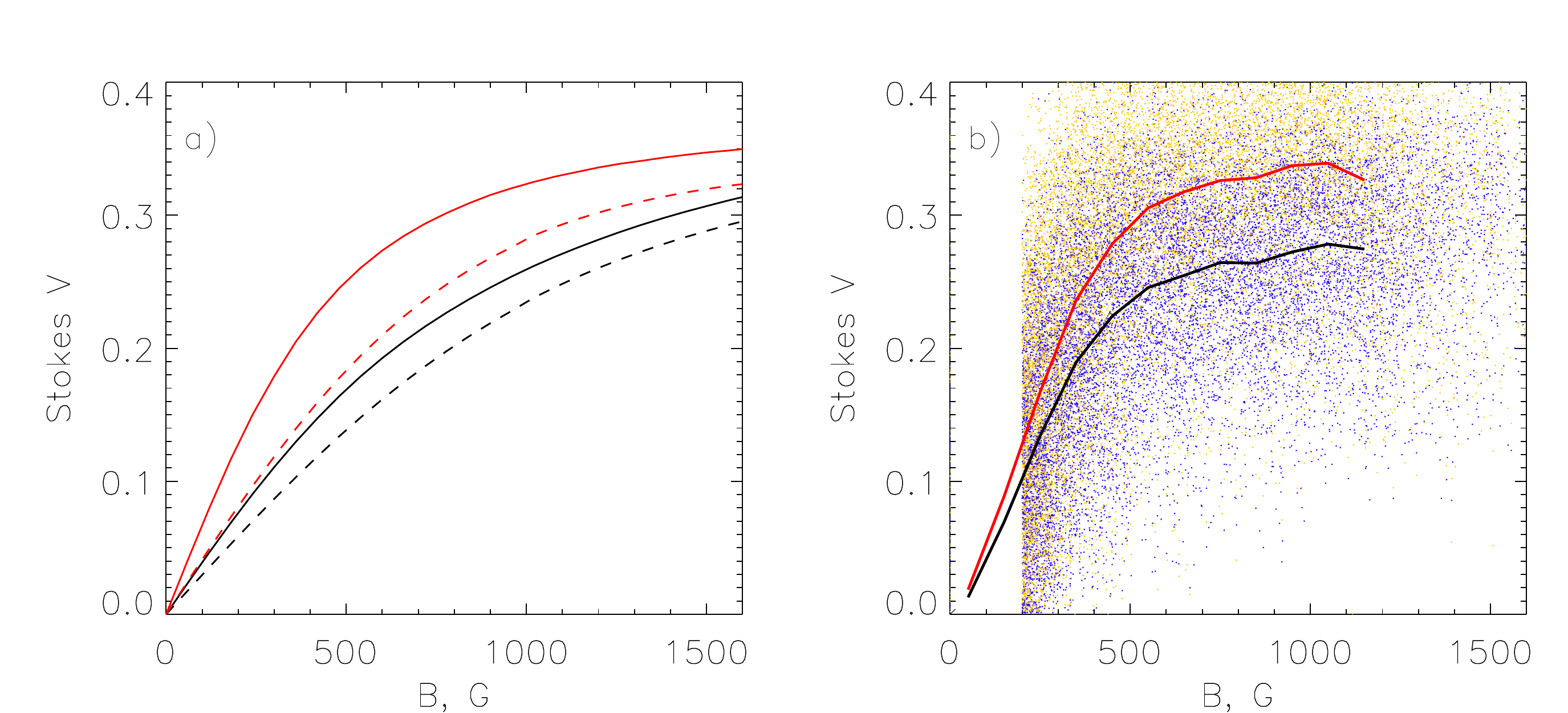}}
\caption{Panel (a): calibration curves for the considered Fe~I lines. Black and red lines are for the 6301.5 and 6302.5~$\ang$ lines, respectively. Solid and dashed lines are for line width of 0.12 and 0.23~$\ang$, respectively. 
Panel (b): Stokes V amplitudes taken from the MURAM-NICOLE simulations. Blue and orange dots are for the 6301.5 and 6302.5~$\ang$ lines, respectively. Black and red lines show sliding averages for 6301.5 and 6302.5~$\ang$ lines, respectively.}
\label{f-callibrflux}
\end{figure*}
\begin{figure*}[ht!]    
\centerline{\includegraphics[width=0.7\textwidth,clip=]{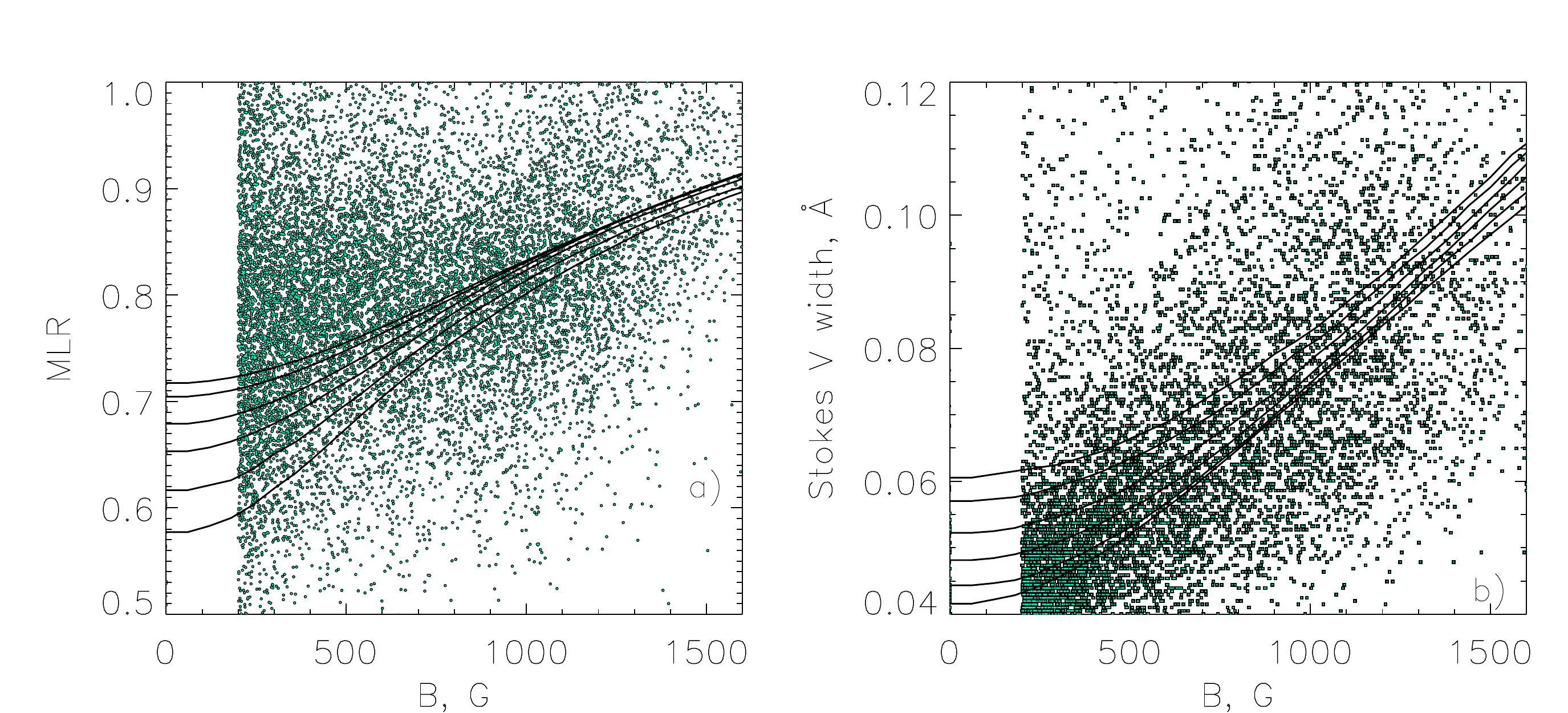}}
\caption{Panel (a): calibration curves for MLR. Panel(b): calibration curves for Stokes V widths of the 6301.5~$\ang$ line. 
In both panels different lines correspond to different widths of $I\pm V$ profile of the 6301.5~$\ang$ line: lowest lines corresponds to 0.12~$\ang$ and highest line corresponds to 0.23~$\ang$. Green dots show MLR values (left panel) and Stokes V widths (right panel) taken from the MURAM-NICOLE simulations. \mrk{Values below 100~G are ignored.}}
\label{f-callibrmagn}
\end{figure*}

3D models of the photospheric magneto-convection have been developed using the radiative MHD code MURAM \cite[see][]{sche03,shee04,voge05}. Simulations have been performed using a box with a uniform grid of 960$\times$960$\times$400 elements with dimensions 6000$\times$6000$\times$2000~km. The upper boundary of the domain corresponds to the temperature minimum level. The simulation starts with a well-developed low-resolution ($240\times 240 \times 160$ grid cells) non-magnetic photospheric convection model. At this point, a uniform vertical magnetic field is added. The model is then run for about 5 convective timescales (corresponding to 50 minutes of physical time). After this initial stage the model is interpolated onto a grid with doubled resolution and run for 10 minutes of physical time. The process is repeated twice to reach the resolution of $6.25\times 6.25\times 5~\mathrm{km}^3$ per grid cell.

\begin{figure*}[ht!]    
\centerline{\includegraphics[width=0.9\textwidth,clip=]{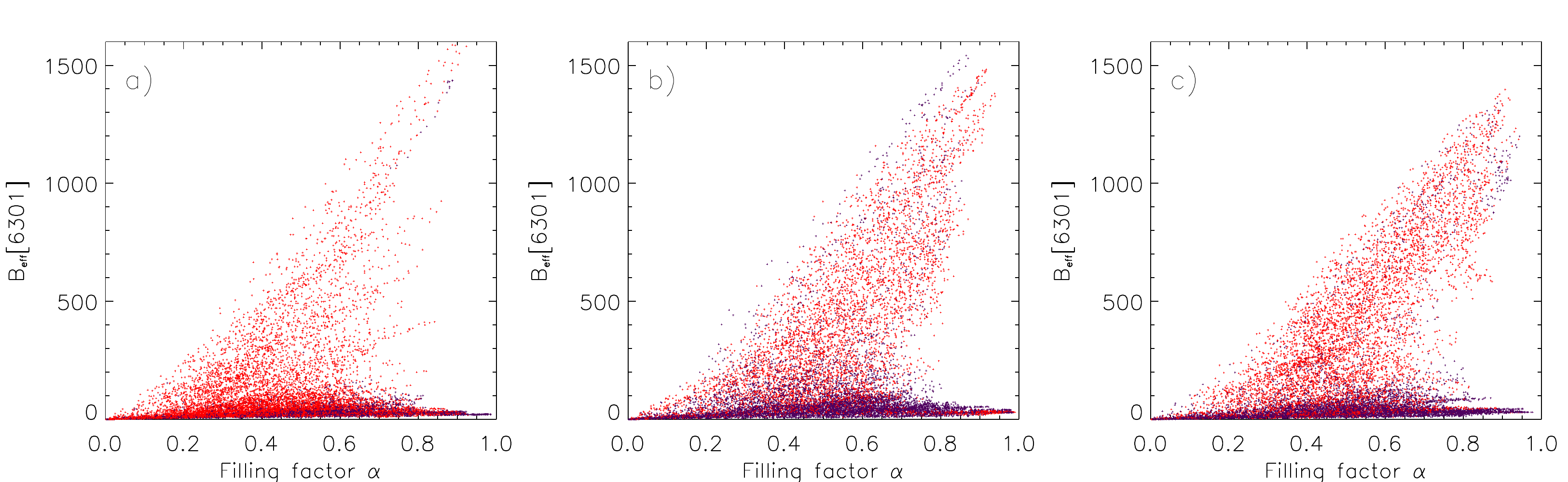}}
\caption{Magnetic field values corresponding to low spatial resolution against the filling factor values in snapshots A, B and C (panels a, b and c, respectively). Red and blue dots correspond to positive and negative LOS velocities (positive and negative Doppler shifts).}
\label{f-alphaflux}
\end{figure*}
\begin{figure*}[ht!]    
\centerline{\includegraphics[width=0.9\textwidth,clip=]{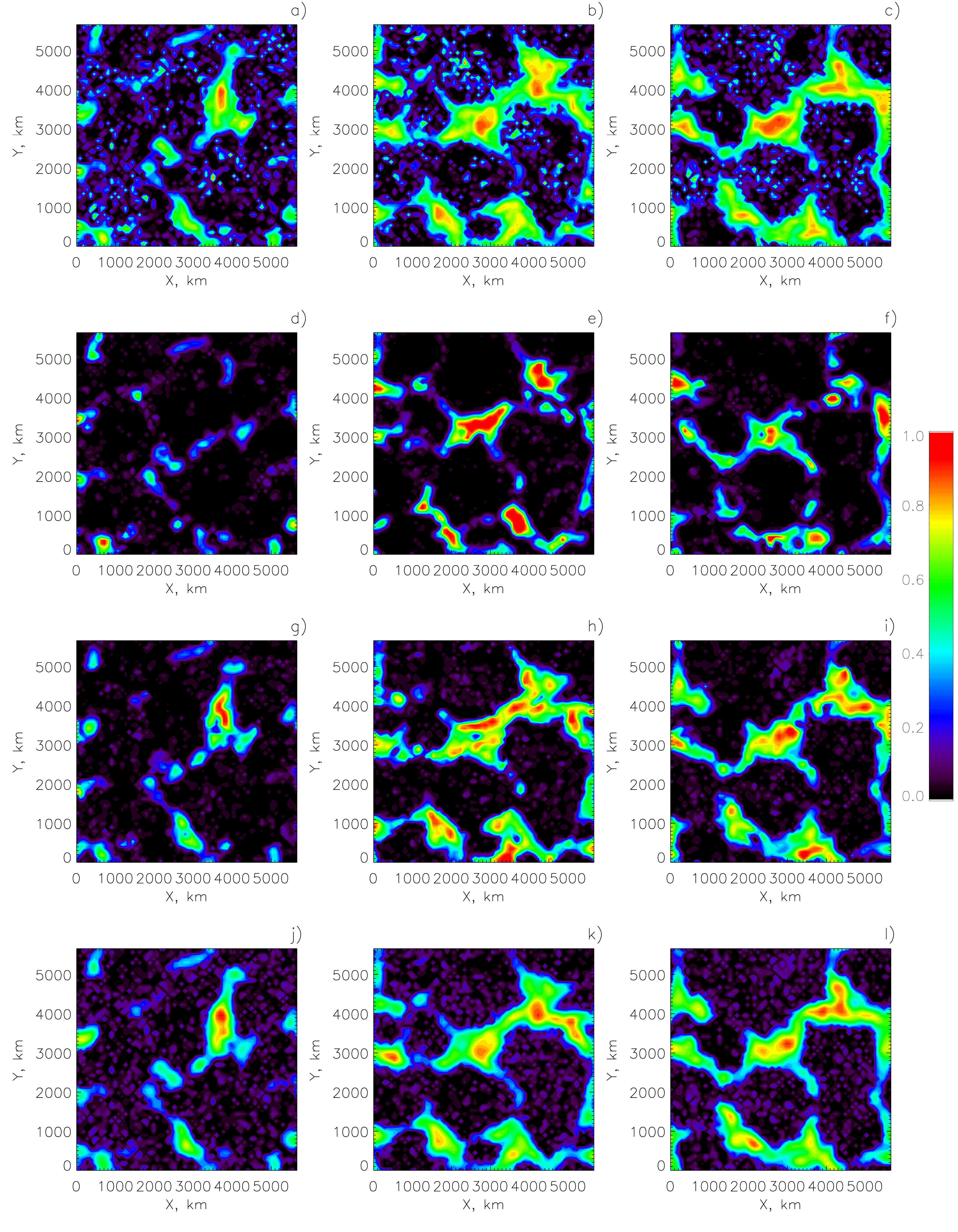}}
\caption{Magnetic filling factor $\alpha$ maps for snapshots A (left panels), B (middle panels) and C (right panels). Panels a-c show $\alpha$ calculated using actual field (i.e. using formula (1) and MHD data), panels d-f, g-i and j-l shows $\alpha$ calculated using the MLR method (Equation 6), Stokes V width method (Equation 7) and statistical method (Equation 8), respectively. Magnetic filling factor values in pixels with magnetic field $|B|<100$G is ignored.}
\label{f-alphamap}
\end{figure*}
\begin{figure*}[ht!]    
\centerline{\includegraphics[width=0.9\textwidth,clip=]{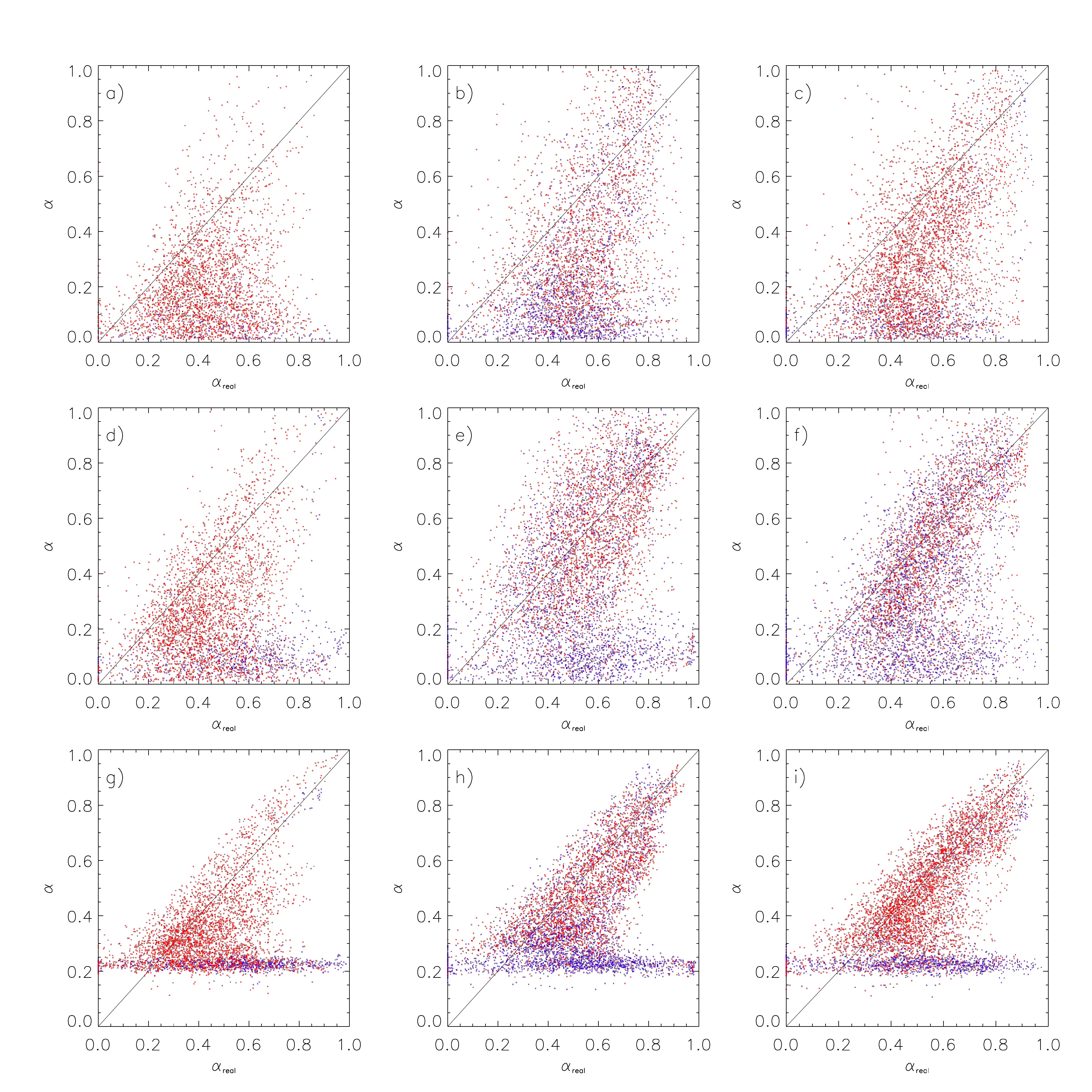}}
\caption{Comparison of estimated values of the magnetic filling factor with real values. Left panels are for snapshot A, middle panels are for snapshot B and right panels are for snapshot C. Panels a-c, d-f and g-i correspond to MLR method, Stokes V width method, and statistical method, respectively. Only pixels with magnetic field $|B|>100$G are shown here.}
\label{f-alphacheck}
\end{figure*}

In this study we use three snapshots from two different models. Snapshot A corresponds to the model with the initial field 100~G, while snapshots B and C correspond to the model with the initial field 200~G. Snapshot B represents a stage when the magnetic structure still shows some changes, while  A and C represent quasi-stationary states: although small-scale elements keep varying, the large scale structure and, most importantly, the magnetic field spectrum do not change. The magnetic field map for snapshot C is shown in Figure~\ref{f-fieldmap}(a).

The evolution of the considered models has been described in detail by \citet{voge05}. After magnetic field is added to the domain, it quickly becomes very non-uniform: convective motions redistribute it so the field becomes concentrated at the boundaries of granulation cells, forming the photospheric network. The field strength is amplified by almost an order-of-magnitude; thus, in the model with initial field of 200~G, when the network is fully developed the maximum field is around 1800--2000~G. Horizontal magnetic field structure appears to be formed of thin walls and larger blobs. The field is non-uniform at scales as little as 10~km (close to horizontal resolution of the computational grid, 6.25~km). Importantly, this scale-length is well below typical resolution of existing optical solar telescopes (50-100~km). 

Histograms of magnetic field values for all three model snapshots are shown in Figure~\ref{f-histomag}. It shows that magnetic field distribution is not ``turbulent'': its spectrum is not a monotonically decreasing, nearly power-law function. Instead, snapshots B and C reveal two peaks: one at $B=0$ and another around $B\approx 1$~kG. (It should be noted that \citet{voge05} show that the weak field peak is located around few tens of Gauss, i.e. the magnetic field is non-zero everywhere in the domain, although most of the model photosphere is permeated by very weak field.) In other words, the intrinsic photospheric magnetic field is represented by two distinct populations: strong small-scale elements found mostly in the network boundaries embedded into ambient weak field. As far as the weaker initial field model is concerned, its field histogram does not show two clear peaks. However, it appears that it still has a multi-component field structure, this will be discussed in Section~\ref{s-lowalpha}.

A two-component field model has been used as a simplification necessary to reduce the parametric space in many studies \cite[see e.g.][]{golo13}. Furthermore, most inversion algorithms (SIR, NICOLE) characterise local magnetic field only by two parameters, strength and filling factor, effectively, assuming the two-component small-scale field structure. However, these MHD models of magneto-convection show that, indeed, the intrinsic photospheric magnetic field has at least two distinct component: strong kiloGauss field mixed with ambient field $B < 50-100$~G. The data from MHD simulations of magneto-convection is used to calculate corresponding I and V stokes profiles of Fe~I 6301.51 and 6302.49$\ang$ lines. $I(x,y,\lambda)$ and $V(x,y,\lambda)$ cubes are calculated using radiative transfer code NICOLE \citep{sone00,sone11}, ignoring possible non-LTE effects and horizontal scattering (in other words, radiative transfer is fully one-dimensional). The wavelength interval is 6301--6303~$\ang$ with the step 5~$m\ang$. The excitation potentials $V$=3.65~eV and $V$=3.686~eV, $\log(gf)$=-0.59 and $\log(gf)$=-1.13 for 6301~$\ang$ and 6302~$\ang$ lines, respectively.

Now, let us discuss the parameters required for quantitative description of the magnetic field inhomogeneity. Vertical inhomogeneity of magnetic field in simplest case can be described using the average field gradient $dB_z/dz$. Horizontal inhomogeneity is normally described using magnetic filling factor $\alpha$. It is assumed that the field consists of two components -- strong field and background field -- and $\alpha$ is the fraction of the surface area penetrated by the strong field. However, it is not obvious how to determine $\alpha$ in case of arbitrary magnetic field strength spectrum.

A possible definition of the filling factor can be based on the fraction of the area occupied by 50\% of the flux carried by strongest field. Assume that the magnetic field distribution within area $S_\mathrm{tot}$ is $dS/dB=f(B)$, i.e. 
$S_\mathrm{tot} = \int \limits_0^\infty f(B) dB$, and the total flux is 
$\Phi_\mathrm{tot} = \int \limits_0^\infty B f(B) dB$. 
There is a value of magnetic field $B_1$ such that
$\frac 12 \Phi_\mathrm{tot} = \int \limits_{B_1}^\infty B f(B) dB$. The corresponding area 
$S_\mathrm{strong} = \int \limits_{B_1}^\infty f(B) dB$ will be equal $\frac 12 S_\mathrm{tot}$ if the magnetic field is uniform ($f(B) = \delta(B-B_1)$); otherwise $S_\mathrm{strong} < \frac 12 S_\mathrm{tot}$ and it will be smaller, the more inhomogeneous magnetic field. Therefore, the filling factor can be introduced as

\begin{equation}
\alpha = 2 \frac{\int \limits_{B_1}^\infty f(B) dB}{\int \limits_{0}^\infty f(B) dB}, 
\end{equation}
where $B_1$ is the field value satisfying
\begin{equation}
\int \limits_{0}^\infty B f(B) dB = 2 \int \limits_{B_1}^\infty B f(B) dB.
\end{equation}
It is easy to show that when magnetic field is two-component with the strong component $B_1$ and zero background field, the formula for the filling factor would reduce to 
\[
\alpha = \frac{\Phi_\mathrm{tot}}{B_1 S_\mathrm{tot}}.
\]

\section{Calibration curves obtained with analytical field configurations}\label{s-callibr}  

There are several parameters that can be potentially used as proxies of the magnetic flux, magnetic field strength and filling factor. In the presence of horizontal field inhomogeneity, a Stokes profile is a convolution of the field distribution $dS/dB$ and corresponding Stokes profiles ($I(B;\lambda)$ or $V(B;\lambda)$). If the field is two-component with nearly zero ambient field, the resulting Stokes profiles can be expressed as a simple linear combinations: 

\begin{eqnarray}
I(\lambda) &=& (1-\alpha) I_\mathrm{amb}(\lambda) + \alpha I_\mathrm{str}(\lambda),\\
V(\lambda) &=& \alpha V_\mathrm{str}(\lambda)
\end{eqnarray}
Hence, the amplitudes of Stokes V will be reduced by the filling factor $\alpha$, while the ratio of Stokes V peak amplitudes as well as the peak widths should remain the same. In other words, if the magnetic field has two-components with nearly zero ambient field, then the average field is $B_\mathrm{eff} = \alpha B_\mathrm{real}$, the measured Stokes V amplitudes would represent the average field $B_\mathrm{eff}$. At the same time, the MLR $V[6301] / V[6302]$ should not be affected by $\alpha$ and would represent the real magnetic field value. Similarly, Stokes V widths $W_\mathrm{V}$, which also depend on the magnetic field, are not affected by the filling factor and should represent the real field value. This is demonstrated by Figures~\ref{f-histomag}-\ref{f-histowid}. Thus, the histograms of magnetic field measured using $V[6301]$ (Figure~\ref{f-histomag}) significantly change when spatial resolution changes. At the same time, the histograms of the MLR and Stokes V widths (Figures~\ref{f-historat}-\ref{f-histowid}) change only slightly. That is why MLR is commonly used as a proxy for the real field value, although the Stokes V width method approach is almost unknown. These two methods will be compared in Section~\ref{s-low}.

Measuring both the average and real field values, one can estimate the filling factor as 

\begin{equation}
\alpha = B_\mathrm{av} / B_\mathrm{real}.
\end{equation}
In the present study we use both MLR and the Stokes V width to estimate the filling factor.

The MLR approach is a well-studied method which has been extensively used before. Different line pairs, including the 6301-6302 pair have been shown as reliable tools for estimating the intrinsic magnetic field values 
\cite[e.g.][]{sten73,sane88,khco07,smso17}. Analysis of Stokes V profile widths is not a commonly used method, although Stokes V width is used in some solar and stellar magnetic field measurements \citep[e.g.][]{dola09}. In this study, we compare these two methods for the magnetic filling factor estimation.

Calibration curves for $V_\mathrm{eff}[6301](B)$, $V_\mathrm{eff}[6302](B)$, $\left[\frac{V_\mathrm{eff}[6301]}{V_\mathrm{eff}[6302]}\right](B)$ and $W_\mathrm{V}[6301](B)$ have been calculated using NICOLE code \citep{sone00,sone11} for different values of uniform magnetic field from -2000 to 2000~G embedded into the HSRA atmosphere. Line width has been varied by introduction of microturbulence with the amplitude from 0 to 6~km~s$^{-1}$.

Figure~\ref{f-callibrflux} shows the calibration curves $V[6301](B)$ and $V[6302](B)$, as well as actual Stokes V amplitudes obtained from the magneto-convection model. It can be seen that, on average, the actual values of $V$ are similar to the calibration curves, although they seem to saturate faster at $B > 1$~kG. Most importantly, the actual $V$ values show a very significant spread, which cannot be explained by different profile widths. This is not surprising, taking into account strong vertical magnetic field and velocity gradients, as well as the correlation of magnetic field and LOS velocities.

In the present study, we use the calibration curves $V[6301](B)$ to measure the average magnetic field. This is done using calibration function $B_\mathrm{eff} [6301]= G(V[6301], W_{I\pm V}[6301]$, where $G$ is tabulated for a range of input parameters.

Figure~\ref{f-callibrmagn} shows calibration curves of MLR $V_\mathrm{eff}[6301] / V_\mathrm{eff}[6302]$ and Stokes V widths of the 6301~$\ang$ line, $W_\mathrm{V}[6301]$. Different curves in Figure~\ref{f-callibrmagn} correspond to different widths of $I+V$ profile of 6301.5~$\ang$ line. We intentionally use the $I+V$ width, unlike in other studies using Stokes I width; this is because the latter appears to be strongly affected by the magnetic field value, not only by macroturbulence velocity. 
It can be seen that both MLR and Stokes V width can be used for $B_\mathrm{real}$ estimations. The MLR is most efficient (i.e. has highest derivative values) at moderate field values and saturates at 1200-1300~G. At the same time, Stokes V width is efficient above 200-300~G, with $W_\mathrm{V}[6301](B_\mathrm{real})$ being nearly flat at $B_\mathrm{real} <$~250~G. Another benefit of using $W_\mathrm{V}[6301]$ is that it appears to be less affected by the line widening, compared to MLR. Thus, at $B_\mathrm{real} = 1000$~G, change of of $I\pm V$ profile width from 0.15 to 0.25~$\ang$ (corresponding to the macroturbulent velocities of 0 and 4.5~km~s$^{-1}$) results in change of MLR from 0.96 to 1.18, equivalent to the difference of about 600~G at constant width, while Stokes V widths changes from 0.68 to 0.75, equivalent to the difference of less than 200~G.

The MLR and Stokes V width calibration curves are used to estimate the real magnetic field values using calibration functions $B_\mathrm{real} = K(V_\mathrm{eff}[6301]/V_\mathrm{eff}[6302], W_{I\pm V}[6301]$ and $B_\mathrm{real} = L(W_\mathrm{V}[6301], W_{I\pm V}[6301]$, with the functions $K$ and $L$ tabulated for a range of input parameters. Therefore, the filling factor can be estimated either as

\begin{equation}
\alpha = \frac{G(V_\mathrm{eff}[6301])}{K(V_\mathrm{eff}[6301]/V_\mathrm{eff}[6302])}
\end{equation}
or as 
\begin{equation}
\alpha = \frac{ G(V_\mathrm{eff}[6301]) }{ L(W_\mathrm{V}[6301]) }.
\end{equation}

\begin{figure*}[ht!]    
\centerline{\includegraphics[width=0.98\textwidth,clip=]{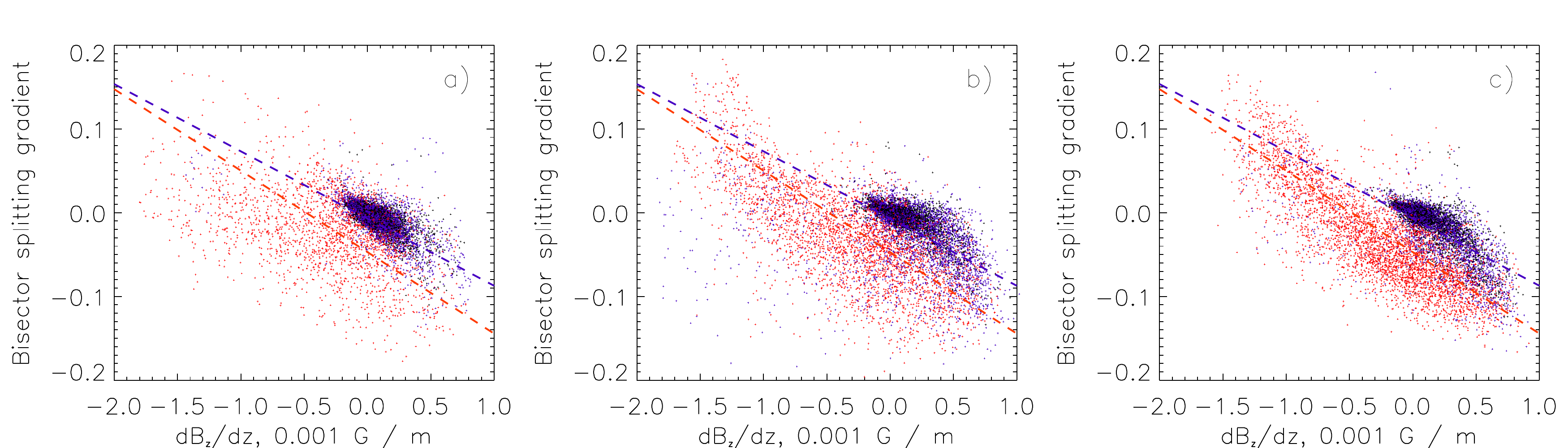}}
\caption{Bisector splitting gradients of $I\pm V$ profiles versus vertical gradients of magnetic field for snapshots A, B and C (panels a, b and c, respectively). Red and blue dots correspond to pixels with positive (downflow) and negative (upflow) Doppler velocities, respectively. Dashed lines show corresponding best linear fits.}
\label{f-callibrgrad}
\end{figure*}

\begin{figure*}[ht!]    
\centerline{\includegraphics[width=0.98\textwidth,clip=]{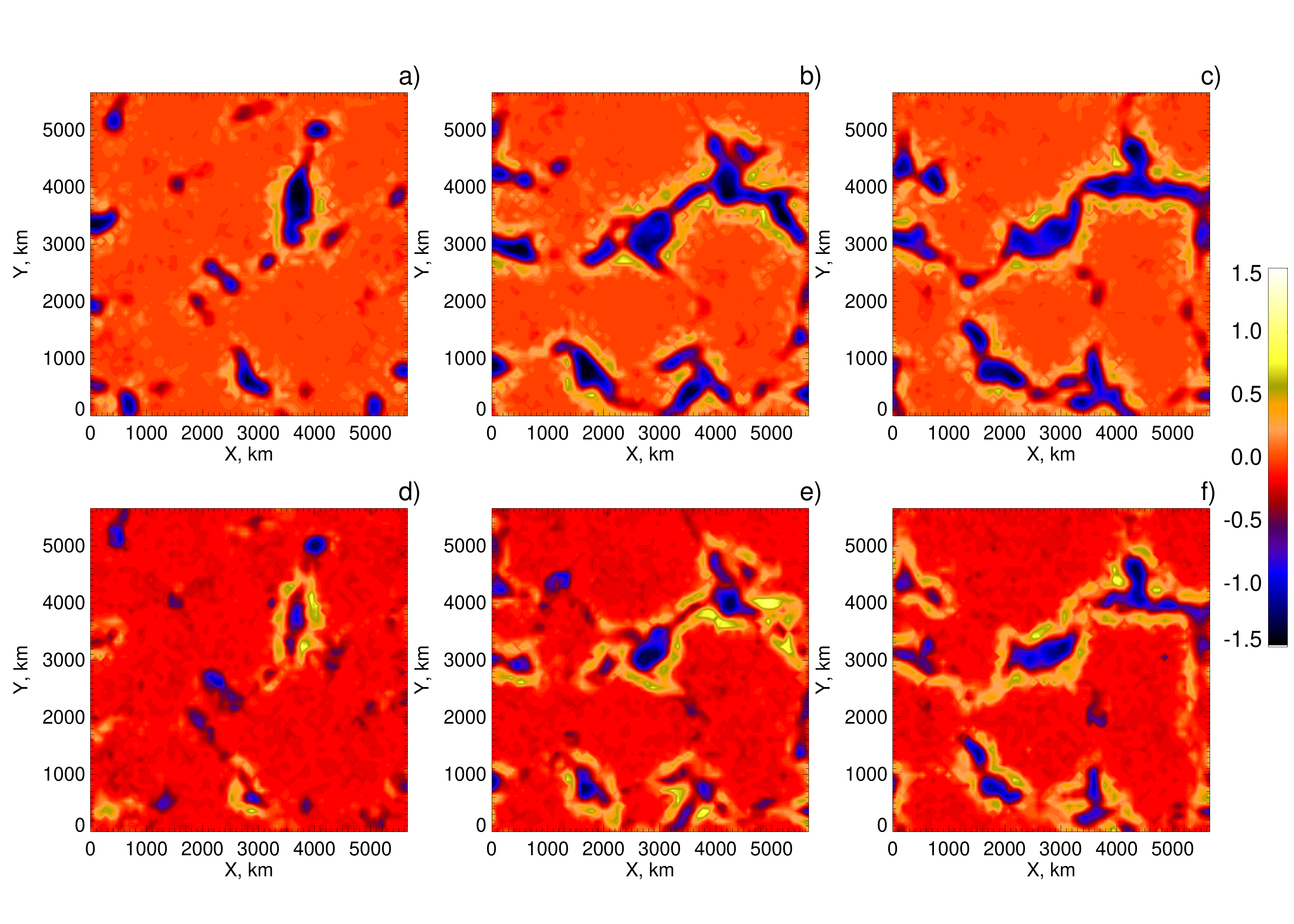}}
\caption{Maps of vertical magnetic field gradients measured directly from the model and reconstructed using BSG. Left, middle and right panels and for snapshots A, B and C, respectively. Panels (a-c) show real gradients (measured from the MHD model), and panels (d-f) show constructed field gradient values.  Colour scale is in units of 10${-3}$~G~m$^{-1}$. Magnetic field gradient values on panels (d-f) are calculated only for pixels with magnetic field $|B|>100$G.}
\label{f-gradimap}
\end{figure*}

\section{Synthetic Stokes profiles calculated with low spatial resolution}\label{s-low}

\subsection{Degraded Stokes profiles}\label{s-lowcalc}

In this section we investigate degraded synthetic I and V Stokes profiles. We recover the filling factor and vertical magnetic field gradient values using degraded synthetic profiles and compare them with those derived using actual magnetic field distributions in the MHD models.

Table 1 compares spatial resolution and the pixel size of Hinode/SOT, Gregor and DKIST/ViSP instruments. The chosen spatial resolution for degraded synthetic data is similar to the three instruments. It is twice higher than in Hinode/SOT and approximately twice lower than in Gregor and DKIST. We do not expect that spatial resolutions differing by factor of 2 would make a substantial difference to the results discussed below and, hence, our analysis should be applicable to the widely used Hinode/SOT, as well as future DKIST/ViSP.

A map of effective magnetic field $B_\mathrm{eff}[6301]$ is shown in Figure~\ref{f-fieldmap}(b). The field distribution is much smoother, it is impossible to see any sharp details on the convective scales.  Unlike high-resolution maps, degraded maps do not show thin magnetic walls. As the results, it looks like the magnetic field consists of numerous blobs in the granulation cell boundaries. Most importantly, the statistical distribution of $B_\mathrm{eff}[6301]$ values is different to that of the intrinsic field. The former does not have kilo-Gauss peaks, it is a smooth, monotonically decreasing function. This can be explained by horizontally nonuniform field structure with the wide range of filling factor variation.

Figure~\ref{f-alphaflux} shows values of $B_\mathrm{eff}$[6301] plotted against $\alpha$, with red and blue colours indicating positive and negative LOS velocities. In snapshots B and C these plots reveal two quite distinct populations. First of them has low magnetic field values ($<$100~G) and mostly positive LOS velocities. This population represents weak magnetic field inside granulation cells. Another population has much stronger magnetic field (up to few kG) and mostly positive Doppler velocities. Hence, this population corresponds to the intergranulation network. The magnetic field values in this population are nearly proportional to the filling factor, $B_\mathrm{eff} \sim \alpha$, although the coefficient of proportionality is different for different snapshots. This effect can be easily explained by the two-component structure of photospheric magnetic field with nearly constant intrinsic field. Indeed, in this case $B_\mathrm{eff} \sim \alpha B$ becomes $B_\mathrm{eff} \sim \mathrm{const} \; \alpha$. Snapshot A, corresponding to the model with weaker initial field, reveals slightly more complicated picture: it seems that the magnetic flux in this model is carried by few magnetic field components, which appear as separate strands in Figure~\ref{f-alphaflux}(a). The fact that the effective magnetic field values are determined primarily by the filling factor is clearly demonstrated by the maps of $\alpha$ (Figure~\ref{f-alphamap}~a-c). Thus, the low-resolution magnetic field map for the snapshot C (Figure~{f-fieldmap}~b) is very similar to the distribution of $\alpha$ for the same snapshot (Figure~\ref{f-alphamap}~c). 

\subsection{Estimation of the filling factor using low-resolution data}\label{s-lowalpha}

Based on the calibration curves discussed in Section 3 and on the low-resolution data analysis discussed in Section 4.1, we can propose three methods to estimate the filling factor assuming two-component magnetic field. Firstly, this can be done using formula (6), with the intrinsic magnetic field estimated using MLR approach. Secondly, this can be done using formula (7) with the intrinsic field estimated using the width of Stokes V peaks. Finally, the filling factor can be estimated statistically, assuming that the intrinsic field is constant within reasonably small area of the photosphere (up to few Mm$^2$). Indeed, for strong magnetic field $>$100~G the filling factor can be approximated as $\alpha \approx \kappa B_\mathrm{eff} + \alpha_0$, with $1/\kappa$ is should approximately be equal to the maximum $B_\mathrm{eff}$ within the sample area. In order to reduce the error in $\kappa$ evaluation, it is calculated as 

\begin{equation}
\alpha \approx \frac {\kappa_0}{B_\mathrm{eff, mean max}} B_\mathrm{eff} + \alpha_0,
\end{equation}
where $B_\mathrm{eff, mean max}$ is the average of top 10\% field values. Best linear fit for the available data is provided by $\alpha_0 = 0.215$ and $\kappa_0 = 0.78$. Maps of the filling factor estimated using different methods are shown in Figure~\ref{f-alphamap}(d-l). 

Generally, all three methods, on average, seem to be yielding values close to the actual ones. However, it is difficult to compare reliability of these methods based on the maps in Figure~\ref{f-alphamap}. Hence, in Figure~\ref{f-alphacheck} ``recovered'' values of the filling factor are plotted against their actual values. \mrk{Because filling factor values are based on the intrinsic field estimations, these plots also demonstrate reliability of $B_\mathrm{real}$ estimations using different methods.} 

It can be seen that using the MLR method (as per Equation 6) yields very substantial spread as well as significantly underestimates filling factor values. In fact, this can also be seen in Figure~\ref{f-alphamap}(d-f). Using the Stokes V widths appears to yield more reliable results. This method slightly underestimates $\alpha$ below 0.5 and slightly overestimates it at higher values, however, it provides much better $\alpha$ estimations compared with using the MLR method. Finally, using the statistical approach (Equation 8) appears to provide best estimations for the filling factor in snapshots B and C: this method seems to give values closest to the $\alpha = \alpha_\mathrm{real}$ line in Figure~\ref{f-alphacheck}(g-i), as well as the smallest spread, compared to two other methods. At the same time, the statistical method does not work well for snapshot A, there are too many pixels with substantially underestimated $\alpha$ values. This can be explained by the multi-component field structure in snapshot A (see Figure~\ref{f-alphaflux}), with the statistical method requiring the intrinsic magnetic field to be nearly constant within the considered photospheric area. 

It should be noted, that it is practically possible to estimate the filling factor values where the magnetic field is significantly above the noise level (approximately 50~G). In Figures~\ref{f-alphamap}-\ref{f-alphacheck} $\alpha$ values are calculated only for B$>$100~G.

\subsection{Estimation of vertical field gradient using low-resolution data}\label{s-lowgrad}

Now, let us discuss the effect of vertical magnetic field inhomogeneity. It is well known that different parts of line profiles are sensitive to different heights in the photosphere: the core is more sensitive to higher layers, close to the temperature minimum, while wings are more sensitive to hotter deeper photospheric layers. Therefore, the vertical distribution of physical parameters in the photosphere can be studied using shapes of the Stokes profile bisectors \cite[e.g.][]{rimm95, trie04, sari02}. Figure~\ref{f-callibrgrad} shows values of BSG plotted against vertical magnetic field gradient separately for positive (network) and negative (cells) LOS velocities. Both snapshots clearly show strong correlation between these values, with $d \Delta \lambda_H / d \delta \lambda$ nearly linearly changing with $dB/dz$. Best linear fits have been calculated using the data from different snapshots added together, however, separately for positive and negative velocities (dashed lines in Figure~\ref{f-callibrgrad}). Regions with positive velocities (photospheric network) show mostly negative field gradients (field becomes stronger with depth), while regions with negative velocities (granulation cells) show predominantly positive field gradients. This also can be seen from maps of $dB/dz$ (Figure~\ref{f-gradimap} a-c). This is in agreement with the widely-accepted model of magnetic field in the solar atmosphere \cite[see e.g. review by][]{wede09}.

The linear fits for the diagrams in Figure~\ref{f-callibrgrad} are
\begin{equation}
\left( \frac{dB}{dz} \right) = k \left( \frac{d \Delta \lambda_H }{ d \delta \lambda} \right) + a,
\end{equation}
where $k$ is $-1.37\times 10^{-4}$ and $-1.83\times 10^{-4}$, and $a$ is $-5.97\times 10^{-5}$ and $-5.30\times 10^{-6}$ for positive and negative velocities, respectively. In this formula $\left( \frac{dB}{dz} \right)$ is in units of G~m$^{-1}$ and $\left( \frac{d \Delta \lambda_H }{ d \delta \lambda} \right)$ is dimensionless ($\ang / \ang$). This simple formula can be used to recover values of magnetic field gradient from using degraded Stokes profiles. Maps of actual and recovered values of magnetic field gradient are compared in Figure~\ref{f-gradimap}. Generally, they are in a good agreement, both qualitatively and quantitatively, although the recovered gradient maps are more noisy and have several patches where negative gradient values are significantly overestimated. \mrk{Most likely, this is because the linear approximation used for gradient evaluation fails below $dB_z/dz < -0.0010$~G/m, which is clearly seen in Figure 12 (b) and (c). Using higher-order polynomials would improve reliability of the method, while making it more computationally expensive, which may be a problem for large field-of-view data analysis.}

\section{Discussion and Summary}\label{s-con}

In this study, we analyse synthetic I and V Stokes profiles of Fe~I 6301.5 and 6302.5~$\ang$ lines derived using a magneto-convection model of the photospheric magnetic field, as well as calibration curves derived for these lines. Based on this analysis, three different methods of estimating the intrinsic magnetic field and the magnetic filling factor in solar photosphere are compared. \mrk{For this pair of lines we show that}:

\begin{itemize}

\item
The Stokes V width method appears to be quite \mrk{reliable for the intrinsic magnetic field and filling factor estimations} above 200-300~G and does not show any saturation up to at least 2~kG. Moreover, Stokes V widths seem to be less sensitive to the line width. Therefore, this method appears to be the best for the $B_\mathrm{real}$ and $\alpha$ estimation using the 6301 \mrk{line}. 

\item
The statistical approach can be very efficient for estimating $\alpha$ values within a small patch of the photosphere when the intrinsic field is likely to be nearly constant. \mrk{Obviously, it can not be applied to large or very inhomogeneous, active photospheric areas.}

\item 
The magnetic line ratio method \mrk{can be used} for intrinsic magnetic field \mrk{and the filling factor} using the 6301~-~6302~$\ang$ pair. However, \mrk{it appears to be the least reliable method because of the formation height difference and saturation}. Furthermore, MLR is more sensitive to the line width (Section~\ref{s-callibr}), and, hence, an error in evaluating the line width would increase the error in derived $B_\mathrm{real}$ values. This, in turn, would translate to even higher error in estimated $B_\mathrm{eff}/B_\mathrm{real}$ values. 

\end{itemize}

However, it should be noted, that MLR approach \mrk{can be very efficient for $B_\mathrm{real}$ and $\alpha$ estimations using other line pairs, which do not saturate in a wider magnetic field range \cite[see][and references therein]{khco07,smso17}. }

Finally, we find that BSG correlates with the LOS gradient of the magnetic field. Linear calibration functions $dB/dz\,(BSG)$ calculated separately for lines with positive and negative Doppler shifts provide quite reliable maps of the photospheric magnetic field gradient.

\begin{acknowledgements}
MG and PKB are funded by Science and Technology Facilities Council (UK), grant ST/P000428/1. Simulations have been performed using DiRAC Data Centric system at Durham University, operated by the Institute for Computational Cosmology on behalf of the STFC DiRAC HPC Facility.
\end{acknowledgements}

\end{document}